\documentclass{ocephys}

\usepackage{tikz}
\usetikzlibrary{decorations.markings}

\usepackage{pgfplots}
\usepgfplotslibrary{fillbetween}
\pgfdeclarelayer{background}
\pgfsetlayers{background, main}
\pgfplotsset{compat=1.18}

\definecolor{blue}{HTML}{3f90da}

\title{Orthogonality of coastal trapped waves}
\author{J\"orn Callies}
\affil{California Institute of Technology, Pasadena, California}
\corr{J\"orn Callies, \href{mailto:jcallies@caltech.edu}{jcallies@caltech.edu} \\[\bigskipamount] This work has been accepted to the Journal of Physical Oceanography. The American Meteorological Society does not guarantee that the copy provided here is an accurate copy of the version of record.}

\newcommand{\dd}[2]{\frac{\d #1}{\d #2}}
\newcommand{\qtext}[1]{\quad \text{#1} \quad}

\begin{document}

\nolinenumbers

\section*{Abstract}

Coastal trapped wave modes are shown to be orthogonal in the sense that they make independent contributions to the energy of the wave field. The hydrostatic Boussinesq dynamics on an $f$-plane, linearized around a state of rest, are formulated as a (generalized) Schr\"odinger equation, which exposes the Hermitian structure of the wave operator that implies the orthogonality of eigenmodes. This formulation, which becomes particularly simple in weak form, is parlayed into a finite-element discretization that preserves the symmetries of the original problem and therefore the energy conservation and orthogonality of modes. The geostrophic momentum approximation, under which the orthogonality of modes has been recognized previously, is reprised and discussed in the present context to emphasize that low-frequency coastal trapped waves are edge waves. Their dynamics are governed by boundary potential-vorticity anomalies, both Bretherton-type contributions familiar from the quasi-geostrophic equations and lateral contributions that are important for steep slopes and coastal walls.

\section{Introduction}

The continental margin of the ocean acts as a waveguide, supporting coastal trapped waves that propagate with the coast to the right in the northern hemisphere \parencite[e.g.,][]{huthnance_coastal_1986}. These trapped waves shape the coastal ocean's response to tidal and wind forcing. They can be tracked for thousands of kilometers as they propagate along continental margins after being excited by local wind events \parencite[e.g.,][]{brink_coastal-trapped_1991}, they are thought to mediate the ocean's response to tidal forcing where this forcing is sub-inertial \parencite[e.g.,][]{huthnance_diurnal_1974,chapman_enhanced_1989,brink_generation_1990,musgrave_influence_2017}, and they shape how open-ocean sea level anomalies are expressed near the coast \parencite{hughes_sea_2019}. The structure of these trapped wave modes is well-characterized, and numerical approaches to their calculation have been known for decades. But a key property of these waves appears to have been overlooked: the wave modes are orthogonal in the sense that they make independent contributions to the energy of the wave field, whether or not the evolution is much slower than the inertial period. The purpose of this paper is to demonstrate this orthogonality and to formulate a simple numerical scheme that preserves this property.

The orthogonality of coastal trapped wave modes has been recognized previously for the slow limit. I revisit these reduced dynamics based on the geostrophic momentum approximation and emphasize the edge wave nature of the modes. In particular, I extend the notion of \citeauthor{bretherton_critical_1966}'s (\citeyear{bretherton_critical_1966}) boundary potential vorticity to allow for steep slopes and coastal walls \parencite[cf.,][]{schneider_boundary_2003}.

The theory of coastal trapped waves in a continuously stratified fluid was developed by \textcite{wang_coastal-trapped_1976,clarke_observational_1977,huthnance_coastal_1978}. All these authors considered the hydrostatic Boussinesq equations on an $f$-plane, linearized around a state of rest with a buoyancy profile that is a function of depth only. The bathymetry has an arbitrary shape in the cross-shore direction but does not vary in the along-shore direction, such that a sinusoidal along-shore structure can be imposed \textit{ab initio}. \textcite{wang_coastal-trapped_1976} further imposed a rigid lid, whereas \textcite{clarke_observational_1977,huthnance_coastal_1978} allowed for free-surface displacements. The dynamics can be posed as an eigenvalue problem, in which the eigenvalue is the frequency~$\omega_n$ and the eigenfunction determines the pressure, velocity, and buoyancy fields of the mode. These authors explored the general properties of the modes, their spatial structures, and the corresponding dispersion curves for a range of idealized and realistic bathymetries and stratification profiles.

\textcite{wang_coastal-trapped_1976} showed that these coastal trapped wave modes are orthogonal in the limit $\omega_n^2 \ll f^2$, expressing the orthogonality both in the form of a boundary integral and as the area integral of the wave energy, with the momentum replaced by the geostrophic momentum (see also Section~\ref{sec:slow}). \textcite{clarke_observational_1977} derived the same orthogonality condition in boundary integral form for waves with $\omega_n^2 \ll f^2$ and an along-shore scale that greatly exceeds the cross-shore scale, although now allowing for free-surface displacements. \textcite{huthnance_coastal_1978} restated the orthogonality condition for $\omega_n^2 \ll f^2$ and also made the connection to the wave energy, now including the potential energy due to free-surface displacements.

The question of orthogonality was raised again recently by \textcite{musgrave_energy_2019}. Seeking to understand the energetics of sub-inertial tides, she pointed out that coastal trapped wave modes make independent contributions to the along-shore energy flux in the slow limit only. \textcite{kelly_coastal_2022} showed that such independence can be achieved beyond the slow limit by a redefinition of the wave modes as solutions to an eigenvalue problem that treats the wave frequency as a parameter and the along-shore wavenumber as the eigenvalue \parencite[cf.,][]{webster_numerical_1987,johnson_spectral_2011}. But these redefined modes do not, in general, make independent contributions to the energy itself (see also Section~\ref{sec:examples}).

I show below that the eigenvalue problem that considers the frequency~$\omega_n$ as the eigenvalue is a (generalized) Schr\"odinger equation with a Hermitian operator, whether or not the slow limit is considered. The eigenvalue problem produces modes that are orthogonal in the sense that they make independent contributions to the wave energy. This form of the eigenvalue problem has the further advantage that it generalizes to more complex geometries that are not uniform in the along-shore direction. Quite generally, the response to a tidal or wind forcing can then be understood mode by mode, with no interaction between the modes in the linear equations, even in a complex geometry. For each mode, the response to the forcing depends on the projection of the forcing onto the modal structure and how close the forcing is to the resonant frequency~$\omega_n$ of the mode.

The energy flux, in contrast, loses its significance in more complex geometries. For each mode, the flux is non-divergent once averaged over the period of the mode. In the presence of multiple modes, the flux emerges from the phase relation between the modes. This is how initially localized energy can spread across the domain. I therefore focus on the standard eigenvalue problem arising from the Schr\"odinger equation with the frequency as the eigenvalue.

\section{Orthogonality of modes}
\label{sec:ortho}

\begin{figure}[t]
  \begin{center}
  \begin{tikzpicture}[scale=0.63, transform shape]
    \begin{scope}[thick, font=\small, decoration={markings, mark=at position 0.5 with {\arrow{>}}}]
      \draw[postaction={decorate}, line cap=rect] (0, 0) -- (0, -1) node[pos=0.5, left, outer sep=4pt, align=center] {coastal wall \\[-\smallskipamount] ($x = 0$)};
      \draw[postaction={decorate}, line cap=rect, name path global=bottom, domain=0:8.89, samples=256] plot (\x, {-4 + 3*exp(-\x^2/4)}) node[right] {$\Lambda$};
      \draw[postaction={decorate}, line cap=rect, name path global=surface] (0, 0) -- (8.89, 0) node[right] {$\Gamma$} node[pos=0.5, above, outer sep=4pt] {free surface ($z = 0$)};
      \node[align=center] at (1, -3) {bottom \\[-\smallskipamount] ($z = -h$)};
      \node at (5.03, -1.83) {$\Omega$};
      \draw[->] (-1.25, -4.25) -- (-0.5, -4.25) node[anchor=north west, inner sep=2pt] {$x$};
      \draw[->] (-1.25, -4.25) -- (-1.25, -3.5) node[anchor=south east, inner sep=2pt] {$z$};
    \end{scope}
    \tikzfillbetween[of=bottom and surface, on layer=background]{blue!25}
  \end{tikzpicture}
  \end{center}
  \caption{Sketch of the two-dimensional domain~$\Omega$. The curve~$\Gamma$ marks the surface, and the curve~$\Lambda$ marks the coastal wall (if present) and the bottom. Both curves are oriented away from the origin.}
  \label{fig:sketch}
\end{figure}
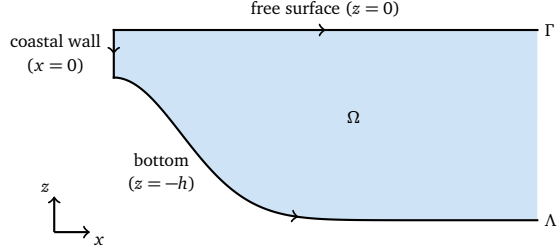

Let us begin with the linearized hydrostatic Boussinesq equations in a domain adjacent to the coast at $x = 0$ and with depth~$h(x)$ (Fig.~\ref{fig:sketch}). Although the discussion that follows is straightforward to generalize, let us consider the specific context of coastal trapped waves by assuming an $f$-plane and no along-shore variations in the depth. The following formulation is entirely equivalent to that employed by \textcite{wang_coastal-trapped_1976,clarke_observational_1977,huthnance_coastal_1978}, among others. Where I depart from most previous work is that I formulate the dynamics as a (generalized) Schr\"odinger equation, which exposes the Hermitian nature of the dynamics and the orthogonality of eigenmodes.

After a Fourier transform in the along-shore direction~$y$ and with the along-shore wavenumber denoted by~$l$, the linearized hydrostatic Boussinesq equations read
\begin{align}
  \partial_t u &= fv - \partial_x p, \\
  \partial_t v &= -fu - il p, \\
  0 &= -\partial_z p + b, \label{eqn:hydbal} \\
  \partial_x u + il v + \partial_z w &= 0, \label{eqn:continuity} \\
  \partial_t b &= -N^2 w, \label{eqn:buoyancy}
\end{align}
where $u$ denotes the cross-shore velocity, $v$~denotes the along-shore velocity, $w$~denotes the vertical velocity, $p$~denotes the kinematic pressure (the dynamic pressure divided by the reference density~$\rho_0$), and $b$~denotes the buoyancy. The stratification~$N^2$ can vary in~$z$. The variables $u$, $v$, $w$, $p$, and~$b$ are the complex Fourier coefficients arising from the along-shore transform. The boundary conditions are no normal flow at the bottom,
\begin{equation}
  w = -u \partial_x h \qtext{at} z = -h(x),
\end{equation}
no flow through the coastal wall (if present), i.e., $u = 0$ at $x = 0$, and a linearized free-surface condition at the top,
\begin{equation}
  g^{-1} \partial_t p = w \qtext{at} z = 0,
\end{equation}
where $g^{-1} p$ is the free-surface elevation. This system conserves energy:
\begin{equation}
  \dd{E}{t} = 0, \quad E = \frac{1}{2} \int_\Omega \left( |u|^2 + |v|^2 + N^{-2} |b|^2 \right) \, \d x \, \d z + \frac{1}{2} \int_\Gamma g^{-1} |p|^2 \, \d x,
  \label{eqn:energy}
\end{equation}
where $\Omega$ denotes the domain in $(x, z)$, and $\Gamma$ denotes the oriented curve along $z = 0$ with orientation away from the origin (Fig.~\ref{fig:sketch}). The total energy consists of kinetic energy, potential energy due to interior buoyancy anomalies, and potential energy due to free-surface displacements. As stated, the energy conservation~\eqref{eqn:energy} holds if the energy density is integrable, which is the case if the flow variables decay sufficiently rapidly as $x \to \infty$ or if the domain is bounded at some finite~$x$.

To better expose the structure of these linear Boussinesq equations, I recast them in operator form:
\begin{equation}
  i M \partial_t \psi = H \psi,
\end{equation}
with the state vector $\psi = (u, v, w, p, b)$ and the operators
\begin{equation}
  M =
  \begin{pmatrix}
    1 \\
    & 1 \\
    & & \hphantom{1} \\
    & & & g^{-1} \delta(z) \\
    & & & & N^{-2}
  \end{pmatrix}, \quad
  H =
  \begin{pmatrix}
    & if & & -i\partial_x \\
    -if & & & l \\
    & & & -i\partial_z & i \\
    -i\partial_x & l & -i\partial_z + i \delta(z) \\
    & & -i
  \end{pmatrix},
\end{equation}
where $\delta$ is the Dirac delta function, and only non-zero entries are shown. The equation takes the form of a Schr\"odinger equation, generalized by the presence of the $M$ operator on the left hand side. The operator $M$~is real, symmetric, and positive semi-definite. The operator $H$ is Hermitian, as can be seen by forming the inner product with $\chi = (\xi, \eta, \zeta, \alpha, \beta)$ and performing integration by parts:
\begin{align}
  \langle \chi, H \psi \rangle &= \int_\Omega \left[ i f (\xi^\dagger v - \eta^\dagger u) - i (\xi^\dagger \partial_x p + \alpha^\dagger \partial_x u) + l (\eta^\dagger p + \alpha^\dagger v) \right. \nonumber \\
  & \qquad \quad \left. - \, i (\zeta^\dagger \partial_z p + \alpha^\dagger \partial_z w) + i (\zeta^\dagger b - \beta^\dagger w) \right] \, \d x \, \d z + \int_\Gamma i \alpha^\dagger w \, \d x \nonumber \\
  &= \int_\Omega \left[ -i f (\eta^\dagger u - \xi^\dagger v) + i (\partial_x \alpha^\dagger u + \partial_x \xi^\dagger p) + l (\alpha^\dagger v + \eta^\dagger p) \right. \nonumber \\
  & \qquad \quad \left. + \, i (\partial_z \alpha^\dagger w + \partial_z \zeta^\dagger p) - i (\beta^\dagger w - \zeta^\dagger b) \right] \, \d x \, \d z - \int_\Gamma i \zeta^\dagger p \, \d x \nonumber \\
  &= \langle H \chi, \psi \rangle,
\end{align}
where $^\dagger$ denotes complex conjugation, and I exploited the no-normal-flow boundary conditions at the bottom and the coastal wall (if present). The inner product is defined as conventional:
\begin{equation*}
  \langle \chi, \psi \rangle = \int_\Omega (\xi^\dagger u + \eta^\dagger v + \zeta^\dagger w + \alpha^\dagger p + \beta^\dagger b) \, \d x \, \d z,
\end{equation*}
such that $\langle \chi, a \psi \rangle = a \langle \chi, \psi \rangle$ and $\langle a \chi, \psi \rangle = a^\dagger \langle \chi, \psi \rangle$ for a complex scalar~$a$.

Let us seek eigenmodes~$\psi_n$ of the operator pair~$(H, M)$ that satisfy
\begin{equation}
  H \psi_n = \omega_n M \psi_n. \label{eqn:eigfull}
\end{equation}
The eigenvalues~$\omega_n$ are real because $H$ is Hermitian and $M$ is positive semi-definite, such that each mode oscillates with $e^{-i \omega_n t}$. Any two eigenmodes $\psi_n$ and $\psi_m$ with distinct frequencies $\omega_n \neq \omega_m$ are $M$-orthogonal, as can be demonstrated by a standard argument:
\begin{multline}
  \omega_n \langle \psi_m, \psi_n \rangle_M = \omega_n \langle \psi_m, M \psi_n \rangle = \langle \psi_m, H \psi_n \rangle \\
  = \langle H \psi_m, \psi_n \rangle = \omega_m \langle M \psi_m, \psi_n \rangle = \omega_m \langle \psi_m, \psi_n \rangle_M, \qtext{so} \langle \psi_m, \psi_n \rangle_M = 0,
\end{multline}
where
\begin{equation}
  \langle \chi, \psi \rangle_M \equiv \langle \chi, M \psi \rangle = \langle M \chi, \psi \rangle.
\end{equation}
For degenerate eigenmodes with $\omega_n = \omega_m$ (such as geostrophic modes), an $M$-orthogonal basis can be found for the subspace spanned by these modes. If the motion is captured by discrete eigenmodes, i.e., if one can write
\begin{equation}
  \psi = \sum_n a_n \psi_n e^{-i \omega_n t},
\end{equation}
and if the eigenmodes are normalized such that $\langle \psi_n, \psi_n \rangle_M = 1$, then the (properly orthogonalized) eigenmodes diagonalize the energy:
\begin{equation}
  E = \frac{1}{2} \langle \psi, \psi \rangle_M = \frac{1}{2} \sum_n |a_n|^2.
\end{equation}
Each mode makes an independent contribution to the energy. There are no cross terms. The amplitudes $a_n$ are determined by projecting an initial condition or forcing onto the modes.

The operator pair $(H, M)$ has the further property that it can be transformed into the pair $(H
', M)$ with a real symmetric~$H'$ by shifting the phases of $u$ and~$w$ by~$\frac{\pi}{2}$. This can be seen by setting $u = i u'$ and $w = i w'$ and obtaining the modified~$H'$ acting on $(u', v, w', p, b)$. It implies that eigenmodes can be chosen to be real in the transformed system, such that the phase must be uniform for each variable in the original system; $u$~and~$w$ are in phase with each other and in quadrature with $v$,~$p$, and~$b$ \parencite[e.g.,][]{huthnance_coastal_1978,kelly_coastal_2022}. This property of coastal trapped waves does not, however, generalize to the three-dimensional case (see Section~\ref{sec:discussion}), and the transformation is not adopted in what follows.

Eigenmodes with distinct along-shore wavenumbers~$l$ should be considered orthogonal because of the orthogonality of their Fourier modes in the along-shore direction. While modes with distinct~$l$ are orthogonal in this three-dimensional sense, they cannot in general be expected to be $M$-orthogonal in the two-dimensional cross-sectional sense discussed above. Consider, for example, two trapped modes with along-shore wavenumbers $l$ and $l + \varepsilon$, denoted by $\psi_n$ and~$\psi_n^{(\varepsilon)}$. As $\varepsilon \to 0$, $\psi_n^{(\varepsilon)} \to \psi_n$ smoothly, such that $\langle \psi_n, \psi_n^{(\varepsilon)} \rangle_M \to \langle \psi_n, \psi_n \rangle_M = 1$ if the modes are normalized accordingly. Therefore, it cannot be the case that $\psi_n$ and~$\psi_n^{(\varepsilon)}$ are $M$-orthogonal for all $\varepsilon > 0$.

It is not guaranteed that the eigenvalue problem~\eqref{eqn:eigfull} is well-posed \parencite[e.g.,][]{sobolev_new_2006,maas_geometric_1995,arnold_topological_1998,buhler_waves_2014}. The well-posedness is most easily assessed by reducing the eigenvalue problem to one equation for~$p_n$, as is standard in the literature (cf., Section~\ref{sec:slow}). Assuming $\omega_n \neq 0$, the eigenvalue problem can be restated as
\begin{equation}
  -(\omega_n^2 - f^2) \partial_z (N^{-2} \partial_z p_n) = -\partial_{xx} p_n + l^2 p_n,
\end{equation}
with corresponding boundary conditions applied. This equation changes from elliptic for $\omega_n^2 < f^2$ to hyperbolic for $\omega_n^2 > f^2$. The elliptic sub-inertial case is well-behaved, and modes are either geostrophic ($\omega_n = 0$) or trapped ($0 < \omega_n^2 < f^2$). The hyperbolic super-inertial case can depend sensitively on the geometry. There are exceptions in semi-infinite domains, in which the super-inertial modes are well-defined and form a continuum of modes that propagate in from infinity, are reflected off the bathymetry, and propagate back out to infinity. But the generic case is ill-posed, and I restrict myself to sub-inertial modes.

\section{Fully prognostic formulation}
\label{sec:prognostic}

The dynamics can be reduced to a purely prognostic set of equations, i.e., one that does not contain any diagnostic equations such as the hydrostatic and continuity equations of the original system. This produces a positive definite operator~$M$ and simplifies the numerical solution discussed in Section~\ref{sec:weakfe}. The reduced formulation will also be the basis of the approximate slow dynamics discussed in Section~\ref{sec:slow}. In this section, however, the dynamics remain unchanged compared to Section~\ref{sec:ortho}.

Dividing the buoyancy equation~\eqref{eqn:buoyancy} by~$N^2$, differentiating it in~$z$, and substituting using the hydrostatic balance~\eqref{eqn:hydbal} and the continuity equation~\eqref{eqn:continuity} eliminate $w$ and~$b$:
\begin{equation}
  \partial_{tz} (N^{-2} \partial_z p) = \partial_x u + i l v.
  \label{eqn:preseq}
\end{equation}
At the free surface,
\begin{equation}
  \partial_t b = -N^2 w = -N^2 g^{-1} \partial_t p, \quad \text{so} \quad \partial_t (N^{-2} \partial_z p + g^{-1} p) = 0 \quad \text{at} \quad z = 0. \label{eqn:surfbc}
\end{equation}
At the bottom,
\begin{equation}
  \partial_t b = -N^2 w = N^2 \partial_x h \, u, \qtext{so} \partial_t (N^{-2} \partial_z p) = \partial_x h \, u \quad \text{at} \quad z = -h.
  \label{eqn:bottombuoy}
\end{equation}
This reduces the dynamics to the three variables $u$, $v$, and~$p$. The eliminated variables $w$ and~$b$ can be recovered using the continuity equation and hydrostatic balance.

One can again write the dynamics in the form
\begin{equation}
  i M \partial_t \psi = H \psi, \label{eqn:alteq}
\end{equation}
with (redefined) $\psi = (u, v, p)$,
\begin{equation}
  M = 
  \begin{pmatrix}
    1 \\
    & 1 \\
    & & -\partial_z (N^{-2} \partial_z \, \cdot \, ) + N^{-2} \delta(z) \partial_z - N^{-2} \delta(z+h) \partial_z + g^{-1} \delta(z)
  \end{pmatrix},
\end{equation}
and
\begin{equation}
  H = 
  \begin{pmatrix}
    & if & -i \partial_x \\
    -if & & l \\
    -i \partial_x - i \partial_x h \, \delta(z+h) & l
  \end{pmatrix}.
\end{equation}
An integration by parts in~$z$ demonstrates that $\frac{1}{2} \langle \psi, \psi \rangle_M$ still produces the correct energy. The Dirac delta function term in~$H$ is required to make it Hermitian, as can be seen by applying the Leibniz integral rule. The operator~$M$ is now positive definite---or positive semi-definite if a rigid lid is applied by omitting the $g^{-1} \delta(z)$ term. Again, the boundary terms are crucial; \textcite{kelly_coastal_2022} pointed out that~$M$ without the boundary terms is not symmetric.

\section{Weak formulation and finite-element approximation}
\label{sec:weakfe}

Perhaps more naturally, the equations may be written in weak form \parencite[e.g.,][]{salsa_partial_2022}. Denoting the test functions by $\chi = (\xi, \eta, \alpha)$, the weak form of~\eqref{eqn:alteq} is
\begin{equation}
  i \langle \chi, \partial_t \psi \rangle_M = \langle \chi, H \psi \rangle,
\end{equation}
or, after some integration by parts,
\begin{multline}
  i \int_\Omega (\xi^\dagger \partial_t u + \eta^\dagger \partial_t v + N^{-2} \partial_z \alpha^\dagger \partial_{tz} p) \, \d x \, \d z + i \int_\Gamma g^{-1} \alpha^\dagger \partial_t p \, \d x \\
  = \int_\Omega [i f (\xi^\dagger v - \eta^\dagger u) + i (\partial_x \alpha^\dagger u - \xi^\dagger \partial_x p) + l (\eta^\dagger p + \alpha^\dagger v)] \, \d x \, \d z. \label{eqn:weak}
\end{multline}
Conservation of energy $E = \frac{1}{2} \langle \psi, \psi \rangle_M$ is now immediate:
\begin{equation}
  \dd{E}{t} = \frac{1}{2} (\langle \psi, \partial_t \psi \rangle_M^\dagger + \langle \psi, \partial_t \psi \rangle_M) = \frac{i}{2} (\langle \psi, H\psi \rangle^\dagger - \langle \psi, H \psi \rangle) = 0,
\end{equation}
given that $H$ is Hermitian, and therefore $\langle \psi, H \psi \rangle$ is real. The weak form~\eqref{eqn:weak} can also be obtained directly from a variational formulation of the problem \parencite[e.g.,][]{salmon_more_2020}.

The weak formulation~\eqref{eqn:weak} can easily be turned into a finite-element approach for numerical solutions \parencite[e.g.,][]{strang_computational_2007}. Restricting the function space to piecewise linear functions on a triangular mesh of the domain~$\Omega$ and denoting by~$\varphi_m$ the basis functions that are equal to~1 at the $m$th~mesh node, 0 at all other nodes, and linear in between, one can represent the solutions in this restricted space by
\begin{equation}
  u = \sum_m u_m \varphi_m, \quad v = \sum_m v_m \varphi_m, \quad p = \sum_m p_m \varphi_m.
\end{equation}
Testing with the basis functions~$\varphi_n$, one obtains the linear algebraic equations
\begin{align}
  i B_{nm} \partial_t u_m &= i f B_{nm} v_m - i G_{mn} p_m, \\
  i B_{nm} \partial_t v_m &= -i f B_{nm} u_m + l B_{nm} p_m, \\
  i (N^{-2} K_{nm} + g^{-1} S_{nm}) \partial_t p_m &= i G_{nm} u_m + l B_{nm} v_m,
\end{align}
where
\begin{gather}
  B_{nm} = \int_\Omega \varphi_n \varphi_m \, \d x \, \d z, \quad
  G_{nm} = \int_\Omega \partial_x \varphi_n \varphi_m \, \d x \, \d z, \\
  K_{nm} = \int_\Omega \partial_z \varphi_n \partial_z \varphi_m \, \d x \, \d z, \quad
  S_{nm} = \int_\Gamma \varphi_n \varphi_m \, \d x.
\end{gather}
One is then left with the finite-dimensional eigenvalue problem
\begin{equation}
  \mat{H} \vec{\psi}_n = \omega_n \mat{M} \vec{\psi}_n,
  \label{eqn:eigmat}
\end{equation}
with a finite-dimensional eigenvector~$\vec{\psi}_n$ and the matrices
\begin{equation}
  \mat{M} = 
  \begin{pmatrix}
    \mat{B} \\
    & \mat{B} \\
    & & N^{-2} \mat{K} + g^{-1} \mat{S}
  \end{pmatrix}, \quad
  \mat{H} =
  \begin{pmatrix}
    & i f \mat{B} & -i \mat{G}^\mathrm{T} \\
    -i f \mat{B} & & l \mat{B} \\
    i \mat{G} & l \mat{B}
  \end{pmatrix}.
\end{equation}
The matrix~$\mat{M}$ is symmetric and positive definite, and the matrix~$\mat{H}$ is Hermitian.

This finite-element discretization thus preserves the symmetry of the operators $M$ and~$H$, such that eigenvalues are still guaranteed to be real and the modes are exactly orthogonal. Results from the discrete system should still be interpreted carefully, however, given the discussion of well-posedness above. In particular, super-inertial eigenvalues and the corresponding eigenmodes cannot in general be expected to converge as the mesh is refined. Sub-inertial modes, in contrast, behave well and exhibit the expected convergence (Appendix~\ref{sec:convergence}). No spurious sub-inertial modes arising from the discretization have been observed, with and without a rigid lid.

\section{Examples}
\label{sec:examples}

To illustrate the approach, exhibit typical mode structures, and confirm agreement with previous calculations, I provide a few examples of sub-inertial coastal trapped wave modes using the finite-element method described in the previous section. I calculate modes for a bumpy bathymetry, a linear bathymetry following \textcite{huthnance_coastal_1978}, and a tanh bathymetry following \textcite{kelly_coastal_2022}.

\begin{figure}[p]
  \centering
  \includegraphics[scale=0.63]{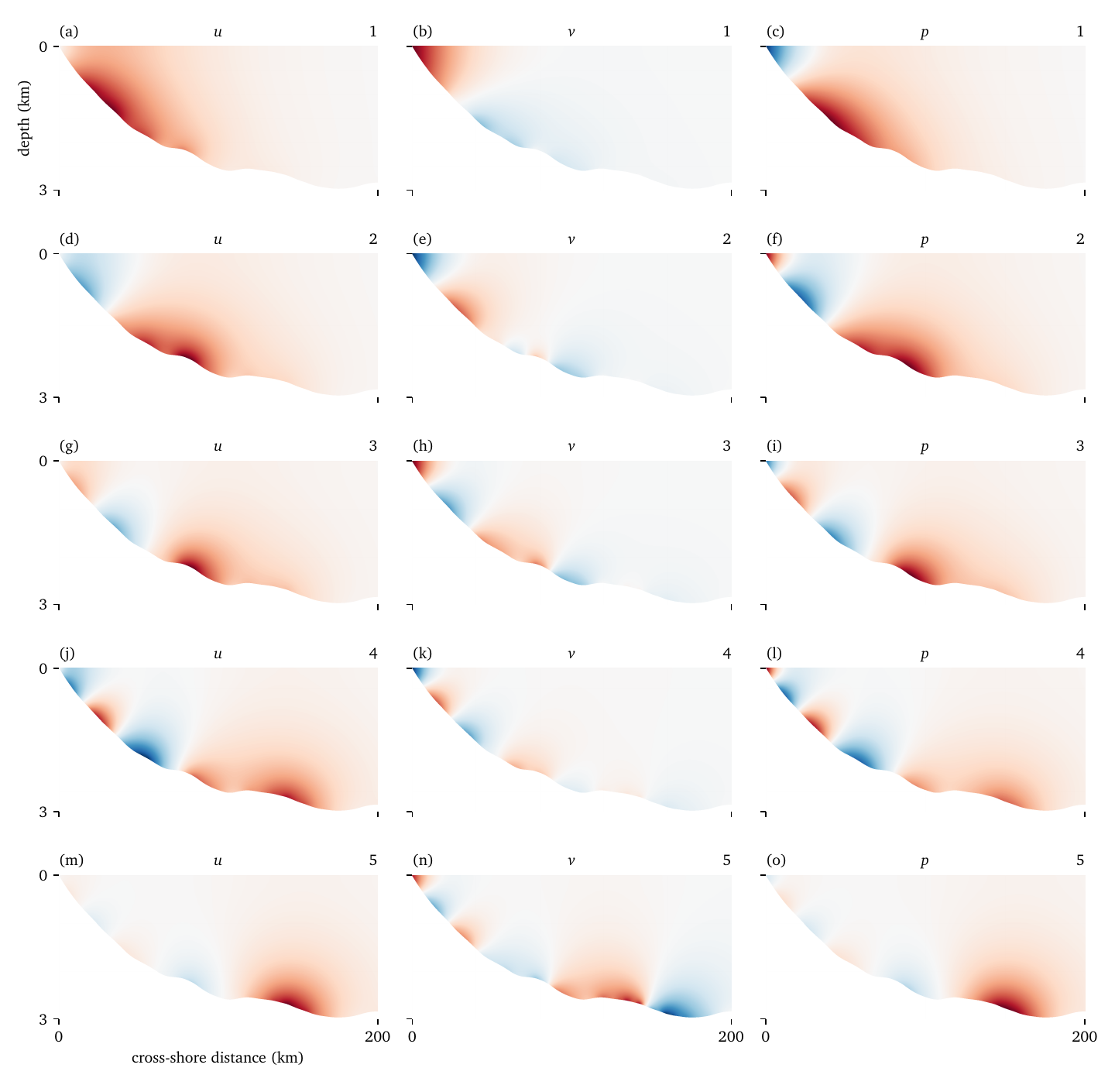}
  \caption{Coastal trapped wave modes 1--5 for a randomly perturbed exponential bathymetry and uniform stratification. These orthogonal modes for an along-shore wavenumber $l = 2\pi / \qty{200}{\kilo\meter}$ have frequencies $\omega_n / f = \numlist{-.52;-.31;-.25;-.19;-.18}$. Only the real part of the cross-shore velocity~$u$ and the imaginary parts of the along-shore velocity~$v$ and pressure~$p$ are shown because the respective other parts vanish. The color scale is normalized to the maximum in each panel. Only the first half of the \qty[number-unit-separator=-]{400}{\kilo\meter}-wide domain is displayed.}
  \label{fig:bumpy}
\end{figure}

First, let us consider a randomly perturbed exponential bathymetry $h(x) = h_0 (1 - e^{-x/d})$ with a decay scale $d = \qty{50}{\kilo\meter}$ (Fig.~\ref{fig:bumpy}). The (additive) random depth perturbations are drawn from a Gaussian PDF using a Mat\'ern covariance function \parencite[e.g.,][]{stein_interpolation_1999} with correlation distance $\rho = \qty{25}{\kilo\meter}$, smoothness parameter~$\nu = \frac{5}{2}$, and standard deviation $\sigma = \min \{\frac{1}{10} h(x), \qty{150}{\meter}\}$. I use a uniform buoyancy frequency $N = \qty{3e-3}{\per\second}$ for simplicity, an inertial frequency~$f = \qty{e-4}{\per\second}$, and an offshore depth of~$h_0 = \qty{3}{\kilo\meter}$.

\begin{figure}[t]
  \centering
  \includegraphics[scale=0.63]{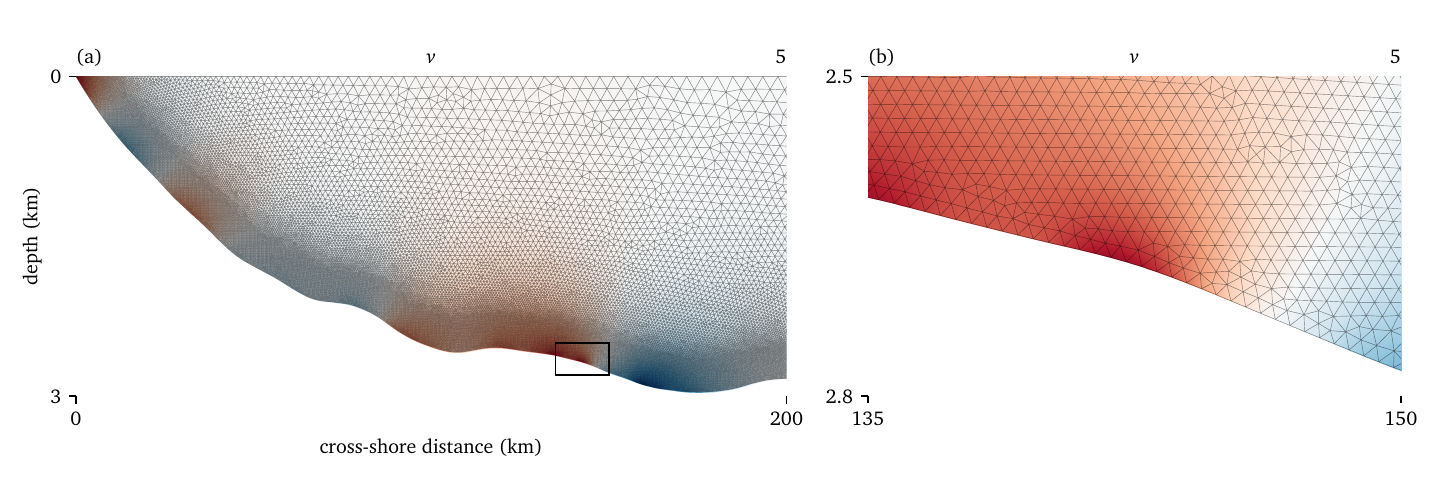}
  \caption{Triangular mesh used in the finite-element calculation of orthogonal coastal trapped wave modes. (left)~Half of the domain and (right)~a zoom-in to the small fraction of the domain indicated by the rectangle in the left panel. The shading shows the along-shore velocity field for mode~5 (see Fig.~\ref{fig:bumpy}n).}
  \label{fig:mesh}
\end{figure}

\begin{table}[b]
  \centering
  \caption{Mesh parameters for the three types of bathymetry considered in this paper. The linear bathymetry is taken from \textcite{huthnance_coastal_1978}, and the tanh bathymetry is taken from \textcite{kelly_coastal_2022}.}
  \label{tab:mesh}
  \begin{tabular}{lcccc}
    \toprule
    bathymetry & domain width & nodes & elements \\
    \midrule
    bumpy & \qty{400}{\kilo\meter} & \num{41907} & \num{82629} \\
    linear & \qty{4000}{\kilo\meter} & \num{15958} & \num{31228} \\
    tanh & \qty{400}{\kilo\meter} & \num{43205} & \num{85157} \\
    \bottomrule
  \end{tabular}
\end{table}

I generate a mesh in non-dimensional space with~$z$ normalized by~$h_0$ and $x$ normalized by $\lambda = N h_0 / f = \qty{90}{\kilo\meter}$, which produces a mesh aspect ratio near the Prandtl ratio $f/N$ (Fig.~\ref{fig:mesh}). I refine the mesh near the bottom, where shorter scales must be resolved. I use a non-dimensional characteristic mesh size of \num{.05} times the distance from the bottom but limit it to a minimum of \num{.005}, corresponding to \qty{15}{\meter} in the vertical and \qty{450}{\meter} in the horizontal. I use a dimensional domain width of \qty{400}{\kilo\meter} to allow modes to decay sufficiently offshore, such that they are not affected appreciably by an artificial offshore wall. This produces a mesh with \num{41907} nodes (Table~\ref{tab:mesh}).

An implicitly restarted Arnoldi iteration applied to $(\mat{H} - \omega_0 \mat{M})^{-1} \mat{M}$ \parencite[e.g.,][]{bai_templates_2000} performs well for the sub-inertial modes. With a shift~$\omega_0$ not too far from the target eigenvalues ($\omega_0 = -\frac{1}{2} f$ works well for the modes shown in Fig.~\ref{fig:bumpy}), the Arnoldi iterations for the first five eigenvalues converge in a handful of iterations and take less than \qty{10}{\second} on a laptop.

The structure of the modes (Fig.~\ref{fig:bumpy}) is consistent with expectations and previous calculations \parencite[e.g.,][]{wang_coastal-trapped_1976,huthnance_coastal_1978,musgrave_energy_2019}. The modes for an along-shore wavenumber~$l = 2\pi / \qty{200}{\kilo\meter}$ are bottom-trapped, and the cross-shore structure decreases in scale with mode number. The number of bottom zero crossings in the pressure modes equals the mode number.

\begin{figure}[p]
  \centering
  \includegraphics[scale=0.63]{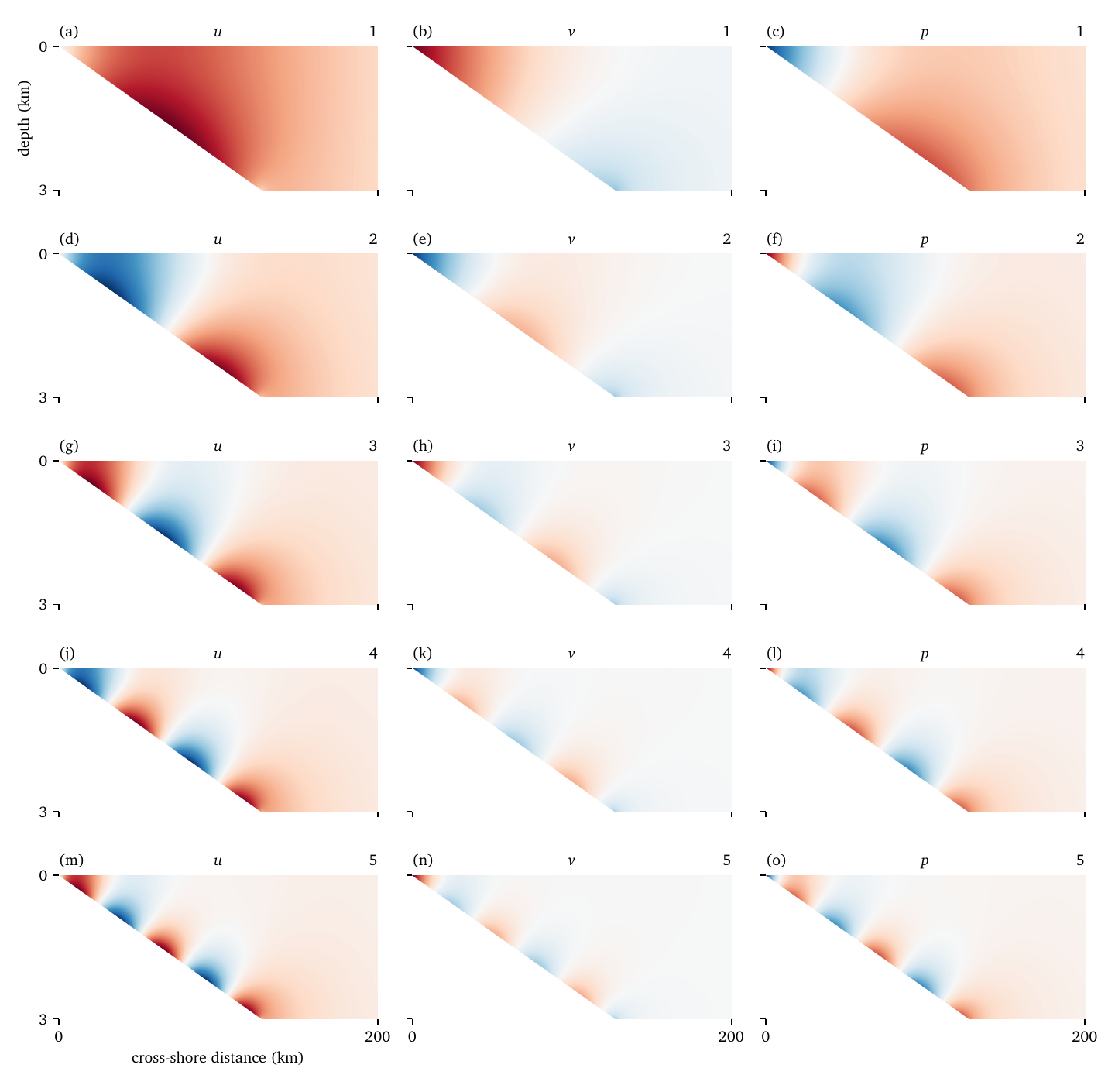}
  \caption{Coastal trapped wave modes 1 through~5 for a bathymetry with a constant slope, uniform stratification, a rigid lid, and a slope Burger number of~$\frac{1}{2}$ \parencite[following][]{huthnance_coastal_1978}. These orthogonal modes for an along-shore wavenumber~$l = \lambda^{-1}$ have frequencies $\omega_n / f = \numlist{-.35;-.18;-.12;-.085;-0.067}$. Only the real part of the cross-shore velocity~$u$ and the imaginary parts of the along-shore velocity~$v$ and pressure~$p$ are shown because the respective other parts vanish. The color scale is normalized to the maximum in each panel. Only the first \qty{200}{\kilo\meter} of the \qty[number-unit-separator=-]{4000}{\kilo\meter}-wide domain is displayed.}
  \label{fig:huthnance}
\end{figure}

Second, I follow \textcite{huthnance_coastal_1978} in choosing a linear bathymetric slope that transitions to a flat bottom offshore (Fig.~\ref{fig:huthnance}). I use the same $N$, $f$, and $h_0$ as before. The slope~$h_0/d$ is chosen such that the slope Burger number is $\lambda^2 / d^2 = \frac{1}{2}$, i.e., the transition to a flat bottom occurs at $x = d = \sqrt{2} \lambda = \qty{127}{\kilo\meter}$. For the reference case, I use an along-shore wavenumber~$l = \lambda^{-1}$. I use a \qty[number-unit-separator=-]{4000}{\kilo\meter}-wide domain to allow for sufficient offshore decay even for small~$l$. I make the mesh size the same function of the distance from the sloping bottom as before, which produces a very coarse mesh far offshore, where the wave modes vary slowly. The number of mesh nodes is smaller than in the previous example because the sloping bathymetry over which the mesh is refined is now confined to $x < \qty{127}{\kilo\meter}$ (Table~\ref{tab:mesh}).

\begin{figure}[t]
  \centering
  \includegraphics[scale=0.63]{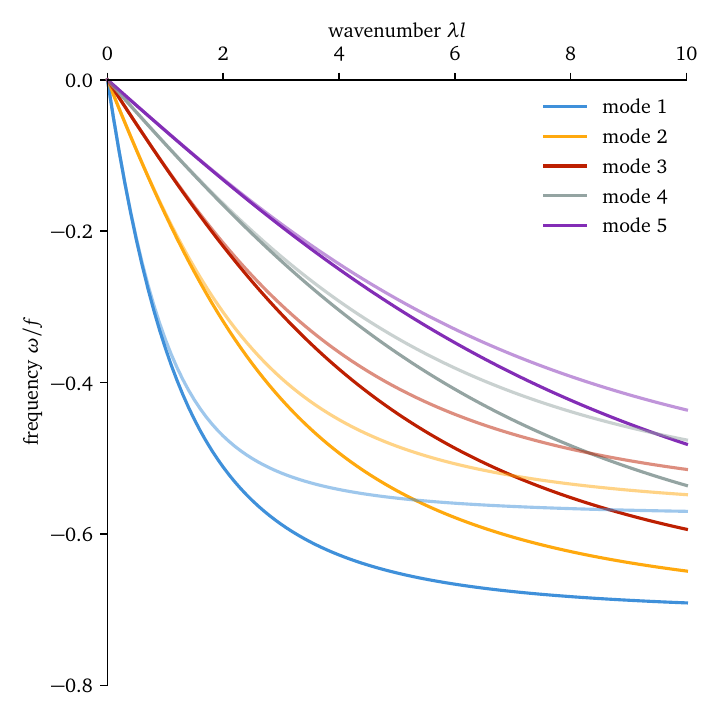}
  \caption{Dispersion curves for coastal trapped modes 1 through~5 for a bathymetry with a constant slope, uniform stratification, a rigid lid, and a slope Burger number of~$\frac{1}{2}$. The solid lines show the eigenvalues of the full problem, and the transparent lines show the eigenvalues of the approximate slow problem (Section~\ref{sec:slow}). The lines for the first two modes can be compared to the corresponding ones in Fig.~8 of \textcite{huthnance_coastal_1978}, although there $l$ is scaled by~$d$ instead of~$\lambda$ and a smaller wavenumber range is displayed.}
  \label{fig:disp}
\end{figure}

The modes again show the expected bottom-trapped structure with an increase in the number of nodes with the mode number (Fig.~\ref{fig:huthnance}). I also vary the along-shore wavenumber to obtain the dispersion curves for the first five modes (Fig.~\ref{fig:disp}), which are consistent with \textcite{huthnance_coastal_1978}. Again, the presently calculated modes do not, in fact, differ from those previously obtained, aside from differences in the numerical approach. Previously, the modes were simply not recognized as orthogonal in the sense discussed here.

\begin{figure}[t]
  \centering
  \includegraphics[scale=0.63]{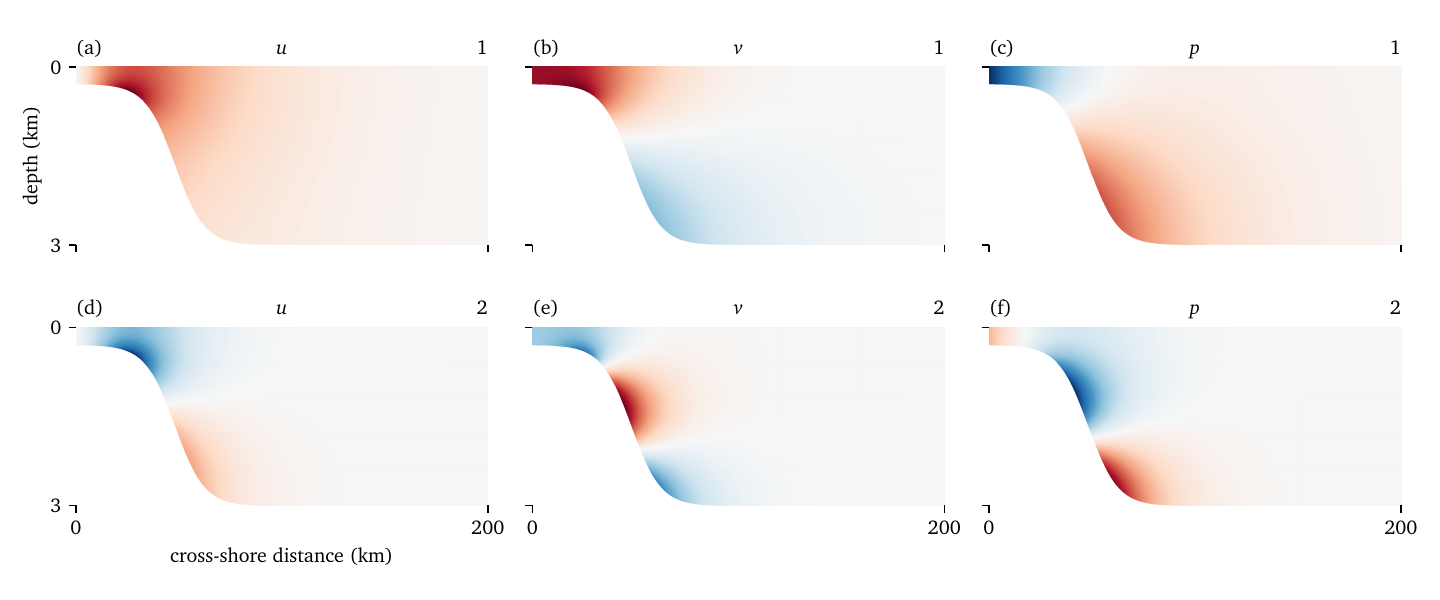}
  \caption{Eigenmodes of~\eqref{eqn:leigen} for the idealized shelf geometry of \textcite{kelly_coastal_2022} and a diurnal tidal period of \qty{24}{\hour}. These modes have along-shore wavelengths of \qtylist{326;128}{\kilo\meter}. Only the real part of the cross-shore velocity~$u$ and the imaginary parts of the along-shore velocity~$v$ and pressure~$p$ are shown because the respective other parts vanish. Only the first half of the \qty[number-unit-separator=-]{400}{\kilo\meter}-wide domain is displayed.}
  \label{fig:kelly}
\end{figure}

The present finite-element discretization can also be used to solve the eigenvalue problem that treats $\omega$ as a parameter and the along-shore wavenumber~$l_n$ as the eigenvalue \parencite[e.g.,][]{webster_numerical_1987,johnson_spectral_2011,kelly_coastal_2022}. In weak form, this eigenvalue problem reads
\begin{equation}
  \langle \chi, \tilde H \tilde \psi_n \rangle = l_n \langle \chi, \tilde \psi_n \rangle_{\tilde M},
\end{equation}
where the operators are defined by
\begin{multline}
  \langle \chi, \tilde H \psi \rangle = \omega \int_\Omega (\xi^\dagger u + \eta^\dagger v + N^{-2} \partial_z \alpha^\dagger \partial_z p) \, \d x \, \d z + \omega \int_\Gamma g^{-1} \alpha^\dagger p \, \d x \\
  - \int_\Omega [i f (\xi^\dagger v - \eta^\dagger u) + i (\partial_x \alpha^\dagger u - \xi^\dagger \partial_x p)] \, \d x \, \d z,
\end{multline}
and
\begin{equation}
  \langle \chi, \psi \rangle_{\tilde M} \equiv \langle \chi, \tilde M \psi \rangle = \int_\Omega (\eta^\dagger p + \alpha^\dagger v) \, \d x \, \d z.
\end{equation}
The discrete eigenvalue problem is then
\begin{equation}
  \tilde{\mat{H}} \tilde{\vec{\psi}}_n = l_n \tilde{\mat{M}} \tilde{\vec{\psi}}_n, \quad \tilde{\mat{M}} = 
  \begin{pmatrix}
    \hphantom{\mat{B}} \\
    & & \mat{B} \\
    & \mat{B}
  \end{pmatrix},
  \quad \tilde{\mat{H}} =
  \begin{pmatrix}
    \omega \mat{B} & -i f \mat{B} & i \mat{G}^\mathrm{T} \\
    i f \mat{B} & \omega \mat{B}  \\
    -i \mat{G} & & \omega (N^{-2} \mat{K} + g^{-1} \mat{S})
  \end{pmatrix}.
  \label{eqn:leigen}
\end{equation}
I discretize the idealized tanh geometry described by \textcite{kelly_coastal_2022} as before, except that I now also refine the mesh near the coastal wall. I prescribe $\omega = 2\pi / \qty{24}{\hour}$ and solve using a shift-and-invert approach with a shift~$l_0 = 2\pi/\qty{200}{\kilo\meter}$. I obtain along-shore wavelengths that are close to those reported by \textcite{kelly_coastal_2022}, $2\pi/l_1 = \qty{326}{\kilo\meter}$ and $2\pi/l_2 = \qty{128}{\kilo\meter}$, and I have ensured that these values are converged. As noted previously, these two modes with distinct~$l_n$ are not $M$-orthogonal. They are $\tilde M$-orthogonal and make independent contributions to the along-shore energy flux $\frac{1}{2} \langle \tilde \psi, \tilde \psi \rangle_{\tilde M}$ \parencite{kelly_coastal_2022}.

\section{Slow motion}
\label{sec:slow}

That coastal trapped wave modes are orthogonal has previously been recognized for the $\omega_n^2 \ll f^2$ limit. I briefly recapitulate this result, treating it in the same framework as used in the rest of the paper, and I apply the same finite-element discretization to it. The discussion emphasizes that sub-inertial coastal trapped waves are best thought of as bottom edge waves.

Let us begin by making the geostrophic momentum approximation \parencite[cf.,][]{hoskins_atmospheric_1972,hoskins_geostrophic_1975}:
\begin{align}
  -i l f^{-1} \partial_t p &= fv - \partial_x p, \\
  f^{-1} \partial_{tx} p &= -fu - i l p, \label{eqn:geostrophicy} \\
  \partial_{tz} (N^{-2} \partial_z p) &= \partial_x u + i l v. \label{eqn:preseqslow}
\end{align}
Cross-differentiating the momentum equations yields the vorticity equation
\begin{equation}
  f^{-2} \partial_t (-\partial_{xx} + l^2) p = \partial_x u + i l v,
\end{equation}
and combining it with~$\eqref{eqn:preseqslow}$ produces
\begin{equation}
  \partial_t [f^{-2} (-\partial_{xx} + l^2) p - \partial_z (N^{-2} \partial_z p)] = 0.
  \label{eqn:intpv}
\end{equation}
This is a conservation statement for interior geostrophic potential vorticity, linearized around a state of rest. If the background stratification is non-uniform, the potential vorticity should be understood as a linearization of geostrophic available potential vorticity \parencite{wagner_available_2015}. In the absence of interior potential-vorticity gradients, as on the $f$-plane and with no mean flow, as considered here, the dynamics are controlled by potential-vorticity anomalies at the boundary \parencite[cf.,][]{schneider_boundary_2003}. At the bottom, adding the buoyancy equation~\eqref{eqn:bottombuoy} and $f^{-1} \partial_x h$ times~\eqref{eqn:geostrophicy} yields
\begin{equation}
  \partial_t (f^{-2} \partial_x h \, \partial_x p + N^{-2} \partial_z p) = -i l f^{-1} \partial_x h \, p \qtext{at} z = -h.
  \label{eqn:surfpv}
\end{equation}
At the surface, as before,
\begin{equation}
  \partial_t (N^{-2} \partial_z p + g^{-1} p) = 0 \qtext{at} z = 0.
  \label{eqn:botpv}
\end{equation}
Equations \eqref{eqn:surfpv} and~\eqref{eqn:botpv} can be read as conservation statements for boundary potential vorticity and combined with the interior potential-vorticity equation~\eqref{eqn:intpv}:
\begin{equation}
  i \partial_t M p = H p,
  \label{eqn:slowdynamics}
\end{equation}
with
\begin{multline}
  M = f^{-2} (-\partial_{xx} + l^2) - \partial_z (N^{-2} \partial_z \, \cdot \,) \\
  + N^{-2} \delta(z) \partial_z - \delta(z+h) (f^{-2} \partial_x h \, \partial_x + N^{-2} \partial_z) + g^{-1} \delta(z),
\end{multline}
and
\begin{equation}
  H = -l f^{-1} \partial_x h \, \delta(z+h).
\end{equation}
The operators $M$ and~$H$ now act on~$p$, the only remaining prognostic variable. With the usual convention, $-f M p$ is the potential vorticity. The interior potential vorticity has no sources or sinks. The bottom boundary potential vorticity consists of a \textcite{bretherton_critical_1966} sheet $f N^{-2} \partial_z p$ and a lateral piece $f^{-1} \partial_x h \, \partial_x p$, which originates analogously from the relative vorticity rather than the stretching term. This boundary potential vorticity is generated by the bottom geostrophic flow $-i l f^{-1} p$ across the topographic slope~$\partial_x h$. The surface boundary potential vorticity consists of a \textcite{bretherton_critical_1966} sheet $-f N^{-2} \partial_z p$ and the sheet $-f g^{-1} p$ that represents the stretching due to free-surface displacements. There are no sources or sinks of surface potential vorticity in this setup. All wave modes are bottom edge waves \parencite[cf.,][]{schmidt_direct_1993}.

If a coastal wall is present, the flow has to additionally satisfy
\begin{equation}
  u = 0, \qtext{so} f^{-1} \partial_{tx} p = - i l p \qtext{at} x = 0.
  \label{eqn:wallbc}
\end{equation}
In this case,
\begin{multline}
  M = f^{-2} (-\partial_{xx} + l^2) - \partial_z (N^{-2} \partial_z \, \cdot \,) \\
  - f^{-2} \delta(x) \partial_x + N^{-2} \delta(z) \partial_z - \delta(z+h) (f^{-2} \partial_x h \, \partial_x + N^{-2} \partial_z) + g^{-1} \delta(z),
\end{multline}
and
\begin{equation}
  H = -l f^{-1} \delta(x) - l f^{-1} \partial_x h \, \delta(z+h).
\end{equation}
The potential vorticity~$-f M p$ now includes the additional term $f^{-1} \partial_x p \delta(x)$ representing the contribution from the coastal wall, and there is an additional source term produced by the geostrophic flow~$-i l f^{-1} p$ across the wall. (The boundary condition at $x = 0$ is $u = 0$, not $-i l f^{-1} p = 0$.) All boundary potential vorticity contributions, except for that due to the free-surface displacement, originate from the boundary-normal component of the vector field $\vec{D} = f^{-2} \partial_x p \, \vec{x} + N^{-2} \partial_z p \, \vec{z}$ \parencite{clarke_observational_1977,schneider_boundary_2003}, although neither of the two gauges \citeauthor{schneider_boundary_2003} discussed produces this~$\vec{D}$ upon approximation. The boundary contributions are needed to absorb the boundary terms that arise from the integration by parts of the interior potential vorticity when forming the energy equation (see below).

\begin{figure}[t]
  \centering
  \includegraphics[scale=0.63]{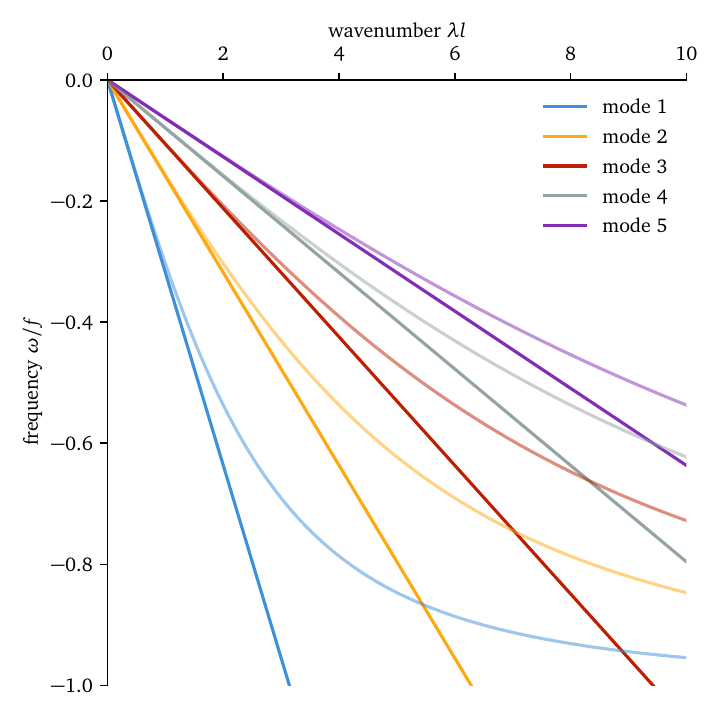}
  \caption{Internal Kelvin wave dispersion curves for the full system (solid lines) and the slow dynamics based on the geostrophic momentum approximation (transparent lines). Note that the lines converge for $\lambda l \to 0$.}
  \label{fig:kelvin}
\end{figure}

The lateral contributions to the boundary potential vorticity, $f^{-1} \partial_x h \, \partial_x p \, \delta(z + h)$ for a sloping bottom and $f^{-1} \partial_x p \, \delta(x)$ for a coastal wall, are absent in a quasi-geostrophic analysis of edge waves \parencite{rhines_edge_1970}. This is because the quasi-geostrophic scaling requires topographic slopes to be much smaller than~$f/N$. The neglect of these lateral terms means that lateral edge waves such as Kelvin waves are not captured. In contrast, the present formulation based on the geostrophic momentum approximation captures the potential vorticity sheet at the coastal wall and produces the following Kelvin wave dispersion relation (imposing a flat bottom at $z = -h_0$):
\begin{equation}
  \frac{\omega_n}{f} = -\frac{l}{\sqrt{l^2 + \lambda_n^{-2}}},
  \label{eqn:slowkelvin}
\end{equation}
where $\lambda_n$ is the $n$th deformation radius \parencite[cf.,][]{kushner_coupled_1998}. For uniform stratification and with a rigid lid, for example, $\lambda_n = \lambda / n \pi$ with $\lambda = N h_0 / f$. For long waves ($l^2 \ll \lambda_n^{-2}$), the condition $\omega_n^2 \ll f^2$ is satisfied, and the approximate dispersion relation~\eqref{eqn:slowkelvin} approaches the Kelvin wave dispersion relation $\omega_n / f = -\lambda_n l$ of the full system (Fig.~\ref{fig:kelvin}).

Whether or not a coastal wall is present, the geostrophic momentum approximation preserves the Hermitian structure of the original system. The system~\eqref{eqn:slowdynamics} conserves the energy
\begin{equation}
  E = \frac{1}{2} \langle p, p \rangle_M = \frac{1}{2} \int_\Omega \left[ f^{-2} \left( |\partial_x p|^2 + l^2 |p|^2 \right) + N^{-2} |\partial_z p|^2 \right] \, \d x \, \d z + \frac{1}{2} \int_\Gamma g^{-1} |p|^2 \, \d x.
  \label{eqn:geoenergy}
\end{equation}
In this expression, geostrophic momentum replaces the momentum that appears in the full energy, as implied by the initial approximation. The approximate system's eigenmodes are $M$-orthogonal, which is to be understood with respect to the redefined~$M$. The eigenmodes make independent contributions to the approximated energy~\eqref{eqn:geoenergy}.

\begin{figure}[t]
  \centering
  \includegraphics[scale=0.63]{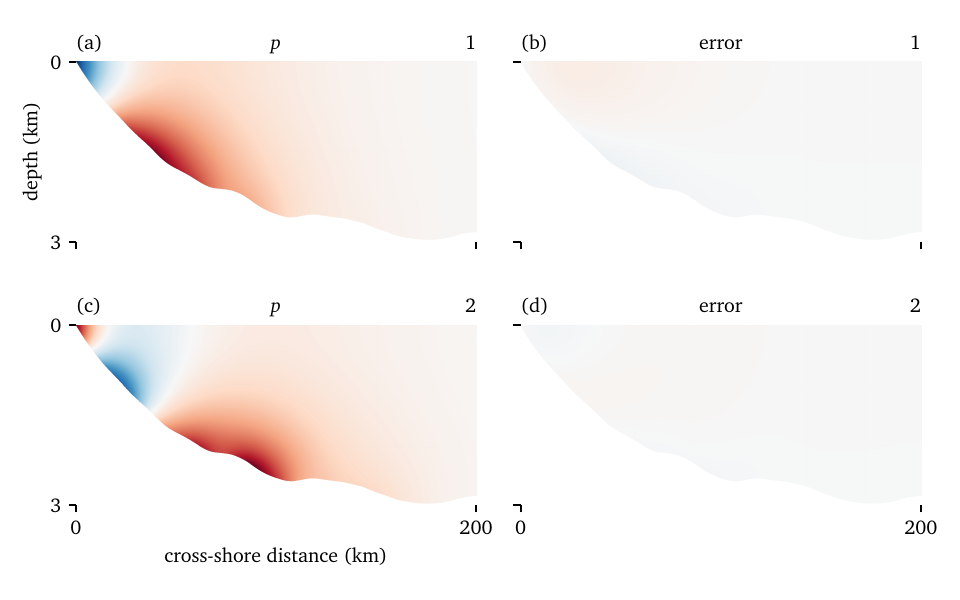}
  \caption{Approximate modes based on the geostrophic momentum approximation for the setup of Fig.~\ref{fig:bumpy}. (left)~The first two approximate modes obtained using~\eqref{eqn:sloweig} and (right)~the differences with the full modes. The numerical error is much smaller than the error due to the approximate dynamics. The color scale is adjusted to the maximum of each mode, and the same color scale is used for the error. The frequencies of the approximate modes are $\omega_n/f = \numlist{-.46;-.30}$, not far from the $\omega_n/f = \numlist{-.52;-.31}$ of the full modes.}
  \label{fig:slow}
\end{figure}

One can again formulate the dynamics in weak form, either by testing the approximated strong form~\eqref{eqn:slowdynamics} or by making the geostrophic momentum approximation in the weak form~\eqref{eqn:weak} of the full dynamics:
\begin{multline}
  i \int_\Omega \left[ f^{-2} (\partial_x \alpha^\dagger \partial_{tx} p + l^2 \alpha^\dagger \partial_t p) + N^{-2} \partial_z \alpha^\dagger \partial_{tz} p \right] \, \d x \, \d z + i \int_\Gamma g^{-1} \alpha^\dagger \partial_t p \, \d x \\
  = \int_\Lambda l f^{-1} \alpha^\dagger p \, \d z,
  \label{eqn:slowweak}
\end{multline}
where $\Lambda$ is the oriented curve that consists of the coastal wall (if present) and the bottom, and its orientation is pointing away from the origin (Fig.~\ref{fig:sketch}). All boundary conditions are enforced naturally, substantially simplifying the weak form compared to the strong form. Compared to the weak form of the full problem, \eqref{eqn:slowweak}~is also simplified substantially because the three terms in the integrand on the right hand side of~\eqref{eqn:weak} combine into a single boundary integral. This reduction to an integral along the coastal wall and bottom again indicates that all waves are edge waves in the slow dynamics. The geostrophic momentum approximation has removed inertia--gravity waves.

The weak form~\eqref{eqn:slowweak} allows writing the orthogonality condition in an alternative form. If $\omega_m \neq 0$, then
\begin{equation}
  \langle p_n, p_m \rangle_M = \langle p_n, M p_m \rangle = \omega_m^{-1} \langle p_n, H p_m \rangle,
\end{equation}
and the orthogonality condition can be expressed as a boundary integral:
\begin{equation}
  \int_\Lambda p_n^\dagger p_m \, \d z = 0
\end{equation}
if $\omega_n \neq \omega_m$ and $l \neq 0$ \parencite[e.g.,][]{wang_coastal-trapped_1976,clarke_observational_1977,huthnance_coastal_1978,brink_energy_1989}. In this commonly quoted form of the orthogonality condition, however, it is not immediately apparent that the modes make independent contributions to the energy. It is the latter property that carries over to the full system in which the geostrophic momentum approximation is not made.

In the approximate slow dynamics, the eigenmodes have the additional property that they make independent contributions not only to the energy $E = \frac{1}{2} \langle p, p \rangle_M$ but also to the along-shore energy flux. Setting $\tilde M = l^{-1} H$, the flux can be written as
\begin{multline}
  F = \frac{1}{2} \langle p, p \rangle_{\tilde M} = \frac{1}{2} \int_\Lambda f^{-1} |p|^2 \, \d z = \frac{1}{2} \int_\Omega f^{-1} \partial_x |p|^2 \, \d x \, \d z \\
  = \frac{1}{2} \int_\Omega f^{-1} (p^\dagger \partial_x p + \partial_x p^\dagger p) \, \d x \, \d z,
\end{multline}
showing that this indeed recovers the pressure work by the geostrophic along-shore flow. The modes make independent contributions to~$F$ because
\begin{equation}
  \langle p_n, p_m \rangle_{\tilde M} = \langle p_n, \tilde M p_m \rangle = \langle p_n, l^{-1} H p_m \rangle = l^{-1} \langle p_n, \omega_m M p_m \rangle = \frac{\omega_m}{l} \langle p_n, p_m \rangle_M,
\end{equation}
such that the $M$-orthogonality of the modes implies $\tilde M$-orthogonality. As discussed by \textcite{musgrave_energy_2019}, who pointed out the simultaneous orthogonality, this result entails advantages in the interpretation of along-shore energy fluxes that do not carry over to the full dynamics. The discussion here suggests that the simultaneous orthogonality is rather fortuitous and cannot be expected to apply to the full system. As mentioned in the introduction, this may not be too important, however, given that the energy flux loses its importance in more general geometries.

As before, one immediately obtains a finite-element approximation of the eigenvalue problem in weak form:
\begin{equation}
  \mat{H} \vec{p}_n = \omega_n \mat{M} \vec{p}_n, \quad \mat{M} = f^{-2} (\mat{L} + l^2 \mat{B}) + N^{-2} \mat{K} + g^{-1} \mat{S}, \quad \mat{H} = l f^{-1} \mat{P},
  \label{eqn:sloweig}
\end{equation}
where now the eigenvector $\vec{p}_n$ contains the nodal values of the pressure field only. The additional matrices are defined as
\begin{equation}
  L_{nm} = \int_\Omega \partial_x \varphi_n \partial_x \varphi_m \, \d x \, \d z, \quad P_{nm} = \int_\Lambda \varphi_n \varphi_m \, \d z.
\end{equation}
Both of these matrices are symmetric, and $\mat{L}$ is positive semi-definite. The eigenvalue problem~\eqref{eqn:sloweig} is somewhat easier to solve than the full problem~\eqref{eqn:eigmat}, given the factor of three reduction in matrix size. The resulting eigenvalues are good approximations to the frequencies of the full modes if $\omega_n^2 \ll f^2$ (Fig.~\ref{fig:disp}). For the example with a bumpy exponential bathymetry (Fig.~\ref{fig:bumpy}), the approximate eigenmodes at $l = 2\pi / \qty{200}{\kilo\meter}$ ($\lambda l = \num{2.8}$) take a very similar shape as the full modes, even for the first two modes for which $\omega_n^2 / f^2$ is not particularly small (Fig.~\ref{fig:slow}).

\section{Discussion}
\label{sec:discussion}

It is straightforward to generalize the dynamics to three dimensions. The eigenvalue problem of the hydrostatic Boussinesq dynamics, written in fully prognostic form, is
\begin{equation}
  \langle \chi, H \psi_n \rangle = \omega_n \langle \chi, \psi_n \rangle_M.
  \label{eqn:weak3d}
\end{equation}
This is again written in weak form and must hold for any test function~$\chi$. The symmetric positive definite~$M$ and the Hermitian~$H$ are defined by
\begin{equation}
  \langle \chi, \psi \rangle_M = \int_\Omega (\xi^\dagger u + \eta^\dagger v + N^{-2} \partial_z \alpha^\dagger \partial_z p) \, \d x \, \d y \, \d z + \int_\Gamma g^{-1} \alpha^\dagger p \, \d x \, \d y
\end{equation}
and
\begin{equation}
  \langle \chi, H \psi \rangle = i \int_\Omega [f (\xi^\dagger v - \eta^\dagger u) + \partial_x \alpha^\dagger u - \xi^\dagger \partial_x p + \partial_y \alpha^\dagger v - \eta^\dagger \partial_y p] \, \d x \, \d y \, \d z,
\end{equation}
where $\Omega$ now denotes the three-dimensional domain, and $\Gamma$ denotes the two-dimensional surface at $z = 0$. The strong form of these operators arises in complete analogy to the two-dimensional case discussed in Section~\ref{sec:prognostic}, now treating both the $x$ and $y$ directions like the $x$~direction there. The eigenvalue problem continues to be well-posed for $\omega_n^2 < f^2$ and yields $M$-orthogonal modes. The weak form~\eqref{eqn:weak3d} is again amenable to finite-element discretization, and a piecewise linear basis on a tetrahedral mesh is expected to behave like the two-dimensional version discussed above. Variations in $N$ and~$f$ can be taken into account by absorbing them into the definition of the associated finite-element matrix.

The slow dynamics can also be generalized to three dimensions. The eigenvalue problem is
\begin{equation}
  \langle \alpha, H p_n \rangle = \omega_n \langle \alpha, p_n \rangle_M
\end{equation}
for all test functions~$\alpha$, where
\begin{equation}
  \langle \alpha, p \rangle_M = \int_\Omega \left[ f^{-2} (\partial_x \alpha^\dagger \partial_x p + \partial_y \alpha^\dagger \partial_y p) + N^{-2} \partial_z \alpha^\dagger \partial_z p \right] \, \d x \, \d y \, \d z + \int_\Gamma g^{-1} \alpha^\dagger p \, \d x \, \d y,
\end{equation}
and
\begin{equation}
  \langle \alpha, H p \rangle = i \int_\Lambda f^{-1} \alpha^\dagger \frac{\partial(p, h)}{\partial(x, y)} \, \d x \, \d y,
\end{equation}
assuming for simplicity that there are no coastal walls and that we remain on an $f$-plane. As in two dimensions, the cross-isobath geostrophic flow provides the only source or sink of potential vorticity.

To extend the modal approach pursued here to super-inertial frequencies, where the modal problem as formulated above is ill-posed in general, one would have to add viscous and diffusion terms to the equations of motion. This would cause energy dissipation and thus spoil the Hermitian nature of the operator. Eigenmodes would not be orthogonal anymore. Calculating left and right eigenvectors and taking advantage of their biorthogonality might alleviate this complication and allow a modal decomposition of arbitrary motion. This framework might be used gainfully to understand the ocean's response to tidal forcing, a response in which mode interactions are restricted to non-linear effects.

\section*{Acknowledgments}

I thank Sam Kelly for feedback on this work and for the encouragement to publish it. All meshes were generated with Gmsh using the Julia API Gmsh.jl. The eigenvalue problems were solved using KrylovKit.jl. I am grateful for funding from the National Aeronautics and Space Administration under grants 80NSSC23K0345 and 80NSSC24K1652.

\section*{Data availability statement}

The source code for the finite-element calculations and mesh generation can be found at \url{https://github.com/joernc/ctw}.

\appendix

\section{Convergence}
\label{sec:convergence}

\begin{figure}[t]
  \centering
  \includegraphics[scale=0.63]{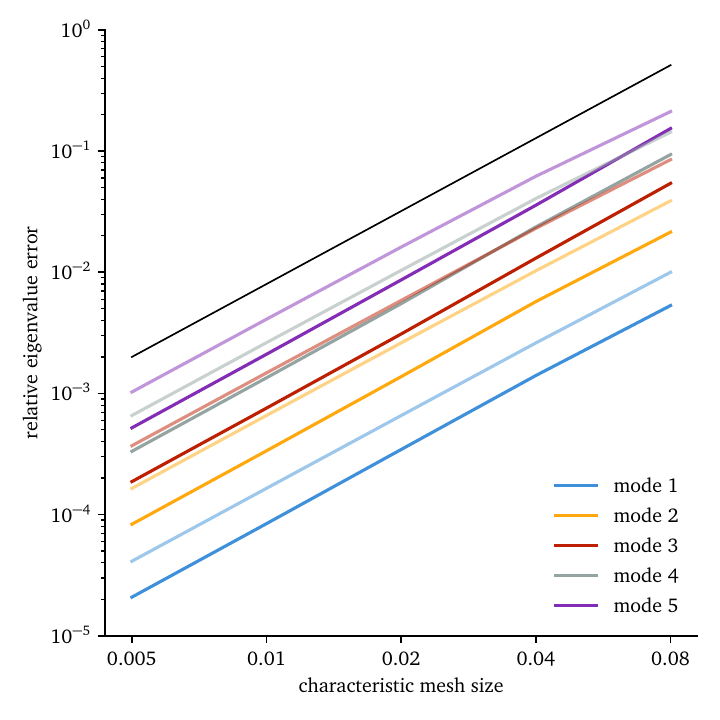}
  \caption{Convergence analysis for the linear finite-element discretization using internal Kelvin wave modes. Shown are the errors of the numerical eigenvalues relative to the exact eigenvalues~$\omega_n/f$ for the first five modes across different characteristic mesh sizes. The non-dimensional mesh sizes are indicated on the horizontal axis. The solid lines show the results for the full dynamics, and the transparent lines show the results for the approximate slow dynamics. The black reference line shows a mesh size squared scaling.}
  \label{fig:convergence}
\end{figure}

I assess the convergence of the finite-element scheme by considering internal Kelvin waves in uniform stratification and with a free surface. With a flat bottom and in a semi-infinite domain, the frequencies of the Kelvin wave modes are $\omega_n/f = -\lambda_n l$. The deformation radii~$\lambda_n$ are determined by finding the roots of
\begin{equation}
  \frac{N h_0}{f \lambda_n} \tan \frac{N h_0}{f \lambda_n} = \frac{N^2 h_0}{g},
\end{equation}
which are approximately
\begin{equation}
  \lambda_n \simeq \frac{N h_0}{n \pi f} \left( 1 - \frac{N^2 h_0}{n^2 \pi^2 g} \right) \qtext{for} n = \text{1, 2, \ldots} \qtext{if} \frac{N^2 h_0}{g} \ll 1.
\end{equation}
I solve the discrete eigenvalue problem~\eqref{eqn:eigmat} using meshes of a \qty[number-unit-separator=-]{1000}{\kilo\meter}-wide domain with characteristic mesh sizes varying from \numrange{0.005}{0.08} in non-dimensional units. The parameters are set to the same values as in the main text: $N = \qty{3e-3}{\per\second}$, $h_0 = \qty{3}{\kilo\meter}$, $f = \qty{e-4}{\per\second}$, and $g = \qty{10}{\meter\per\second\squared}$.

The eigenvalues converge, as expected, with the square of the characteristic mesh size (Fig.~\ref{fig:convergence}). Low modes have a smaller relative error because they vary on a larger scale than high modes. The same convergence rate applies to the approximate slow problem~\eqref{eqn:sloweig}, as assessed by comparing the numerical eigenvalues to the exact eigenvalues~\eqref{eqn:slowkelvin} (Fig.~\ref{fig:convergence}). If a more rapid convergence is desired, a higher-order finite-element scheme can be used.

\end{document}